\documentclass[useAMS,usenatbib]{mn2e}

\usepackage{graphicx}
\usepackage{amsmath}
\usepackage{amssymb}
\usepackage{color}
\usepackage{subfig}

\newcommand\msun{\, \rm M_\odot}
\newcommand\rsun{\, \rm R_\odot}
\newcommand\kms{\, \rm km\,s^{-1}}

\newcommand\kpc{{\, \rm kpc}}
\newcommand\yr{{\, \rm yr}}
\newcommand\gyr{{\, \rm Gyr}}
\newcommand\au{{\, \rm AU}}

\newcommand\rmin{{r_{\rm min}}}
\newcommand\aout{{a_{\rm out}}}
\newcommand\ain{{a_{\rm in}}}

\title[Tidal breakup of triple stars in the GC]{Tidal breakup of triple stars in the Galactic Centre}
\author[G. Fragione and A. Gualandris]{Giacomo Fragione$^{1}$\thanks{E-mail: giacomo.fragione@mail.huji.ac.il} and Alessia Gualandris$^{2}$\\
$^{1}$Racah Institute for Physics, The Hebrew University, Jerusalem 91904, Israel\\
$^{2}$Department of Physics, University of Surrey, Guildford GU2 7XH, United Kingdom}

\begin{document}

\maketitle

\begin{abstract}
The last decade has seen the detection of fast moving stars in the Galactic halo, the so-called hypervelocity stars (HVSs). While the bulk of this population is likely the result of a close encounter between a stellar binary and the supermassive black hole (MBH) in the Galactic Centre (GC), other mechanims may contribute fast stars to the sample. Few observed HVSs show apparent ages which are shorter than the flight time from the GC, thereby making the binary disruption scenario unlikely. These stars may be the result of the breakup of a stellar triple in the GC which led to the ejection of a hypervelocity binary (HVB). If such binary evolves into a blue straggler star due to internal processes after ejection, a rejuvenation is possible that make the star appear younger once detected in the halo. A triple disruption may also be responsible for the presence of HVBs, of which one candidate has now been observed. We present a numerical study of triple disruptions by the MBH in the GC and find that the most likely outcomes are the production of single HVSs and single/binary stars bound to the MBH, while the production of HVBs has a probability $\lesssim 1\%$ regardless of the initial parameters. Assuming a triple fraction of $\approx 10\%$ results in an ejection rate of $\lesssim 1\gyr^{-1}$, insufficient to explain the sample of HVSs with lifetimes shorter than their flight time. We conclude that alternative mechanisms are responsible for the origin of such objects and HVBs in general.
\end{abstract}

\begin{keywords}
Galaxy: centre -- Galaxy: kinematics and dynamics -- stars: kinematics and dynamics -- galaxies: star clusters: general
\end{keywords}

\section{Introduction}

In recent years, a population of stars with extreme radial velocities has been discovered in the Galactic halo, the hypervelocity stars (HVSs).
Predicted by \citet{hills88} as the consequence of close Newtonian encounters of binary stars with the massive black hole (MBH) in the Galactic Centre (GC), the fist HVS was observed by \citet{brw05} in a survey of the Galactic halo, moving with a heliocentric radial velocity of $\sim 700\kms$. With an estimated ejection velocity of $> 1000\kms$, its trajectory is consistent with an origin in the GC \citep{GPZS2005}. 

More than $20$ HVSs have since been confirmed by the Multiple Mirror Telescope survey, with distances between $50$ and $120\kpc$ from the GC and Galactocentric velocities up to $\approx 700\kms$ \citep{brw06,brw12,brw14}.  Yet HVSs remain rare objects and large volume surveys are required for their detection. The ejection rate from the Hills mechanism  in the empty loss cone regime  is $\approx 10^{-6}-10^{-5}\yr^{-1}$\citep{yut03} and  $\approx 10^{-5}-10^{-4}\yr^{-1}$\citep{zhang2013}.
A larger and less biased sample of HVSs is expected from the astrometric European satellite \textit{Gaia} \citep{brw15,mar17}. 

HVSs are important objects in the Galaxy since they can provide an overwhelming amount of information about several astrophysical phenomena in their formation environment and the Galactic potential in which they travel \citep*{yut03,gou03,baum2006,sesa2006,haas2016,fra16,fck17,fragins17}. If they formed in the GC, their distribution in space and velocity can reveal the existence of a secondary MBH, maybe brought by infalling GCs \citep*{fgk17}, while their stellar nature can probe the GC mass function and binary population. In particular, their kinematics can be used to probe the Galactic mass distribution and the triaxiality of the potential 
\citep{gnedin2005,ym2007,gnedin2010,fl2017,rossi2017}. However, many properties of HVSs remain poorly understood, including the ejection mechanism. While an ejection due to a strong dynamical encounter with the MBH in the GC is the favoured model for most of the HVSs, alternative mechanisms have been proposed including encounters with a massive black hole binary in the GC \citep{yut03,baum2006,sesa2006}, encounters in a nearby galaxy \citep{GPZ2007,she2008,bou2017}, tidal interactions of stars clusters with a single or binary MBHs \citep{cap15,acs16,fra16,fck17}, supernova explosions \citep{zub2013,tau2015} and the dynamical evolution of a thin and eccentric disk orbiting the MBH in the GC \citep{sub14,haas2016,sub16}.

A few of the observed HVSs are particularly challenging to reconcile with an origin in the GC given that their travel time from the GC is longer than their evolutionary time \citep[see][for a list of candidates]{per09}. The largest discrepancy between the two timescales is inferred for HE0437-5439, a $9\msun$ B-type main-sequence star moving with a heliocentric radial velocity of about $720\kms$
at a distance of $\sim60\kpc$ \citep{ede05}. Given its proximity to the Large Magellanic Cloud, \citet{ede05} suggest an origin in the LMC, which would require a black hole of at least $10^3\msun$ \citep{GPZ2007}. Such an origin is supported by \citet{bou16}, who showed that the observed clustering of the HVSs in Leo constellation may be explained by an LMC origin. Proper motion measurements for this star are not accurate enough to discriminate between a Galactic and LMC origin \citep{brw15}, and metallicity measurements are inconclusive \citep{przy2008,per09}.
An origin in the GC would only be possible if the star were a blue straggler, i.e. the merger product of a binary ejected at hypervelocity by the MBH. In particular, \citet{per09} suggest that the star was ejected as a hypervelocity binary (HVB) 
as a consequence of a triple star disruption, and later coalesced to form a single rejuvenated HVS, reconciling the discrepancy between its flight time and (apparent) main sequence lifetime. A similar scenario was proposed by \citet{fck17}, who find that $\sim 7\%$ of the HVBs ejected by a compact young star cluster merge originating blue-straggler HVSs.

The hot subdwarf SDSS J121150.27+143716.2 discovered by \citet{tilli2011} has recently been shown to be a HVB \citep{nem16}. Its reconstructed trajectory in the Galactic potential appears inconsistent with an ejection from the GC and an origin in the Galactic halo or an accretion event from a dwarf galaxy seem more likely. However, ejections of HVBs from the GC may be key to explain the short main-sequence lifetimes of some HVSs in the halo.

Here, we study the ejection of HVBs from encounters between triple stars and the MBH in the Galactic Centre by means of scattering experiments. This scenario was considered of negligible importance by \citet{lu2007} but later reconsidered by \citet{per09} who predict significant ejection rates in the case of massive B-type stars.

Stellar multiplicity is an omnipresent outcome of the star-formation process \citep{duc13} and more than $50$\% of stars are thought to have at least one stellar companion \citep{tok14a,tok14b}. \citet{tok14b} show that nearly $13$\% of F- and G-type dwarf stars in the Hipparcos sample live in triple (or higher order) hierarchical systems, while \citet{rid15} find a relatively large abundance of $2$+$2$ quadruples with  
Robo-AO, the first robotic adaptive optics instrument. Among O-stars, about $80\%$ have at least one companion and nearly $25$\% have at least two such companions in their sample \citep{san14}. Using a large high-resolution radial velocity spectroscopic survey of B- and O-type stars, \citet{chi12} find that at least $50$-$80$\% of B- and O-type stars are multiples.
Given the observed high frequency of triple systems, triple disruptions in the GC ought not to be rare.

The paper is organised as follows. In Section 2, we describe the methods and initial conditions used in our numerical experiments. In Section 3 we present the results of the scattering experiments, while 
in Section 4 we discuss the implications of our findings and present our conclusions.

\section{Method}
Let's consider a binary star comprised of equal mass stars of mass $m$ and with semi-major axis $a$ that undergoes a close interaction with a MBH of mass $M$.  The binary is disrupted if it approaches the MBH within a distance equal to the tidal radius 
\begin{equation}
r_t\approx \left(\frac{M}{m}\right)^{1/3} a\ .
\label{eqn:rt}
\end{equation}
The distance of closest approach of the binary centre of mass $\rmin$ can be computed from momentum conservation with respect to the MBH
\begin{equation}
v\,D=\left(\frac{GM}{\rmin}\right)^{1/2} \rmin\ ,
\label{eqn:rmin}
\end{equation}
where $v$ is the transverse speed and $D$ is the initial distance from the MBH. If $\rmin \lesssim r_t$, the binary is disrupted. In general, there are three possible outcomes for binary disruptions: (i) production of an HVS and a S-star; (ii) production of 2 S-stars; (iii) capture of the whole binary. In the case of a triple star, the dynamics becomes more complicated and seven outcome channels are possible, with production of:
\begin{itemize}
\item 1 single HVS and 2 single S-stars (1SH-2SS);
\item 1 single HVS and 1 binary S-star (1SH-1BS);
\item 2 single HVSs and 1 single S-star (2SH-1SS);
\item 1 binary HVS and 1 single S-stars (1BH-1SS);
\item 3 single S-stars (3SS);
\item 1 single S-star and 1 binary S-star (1SS-1BS);
\item 1 triple S-star (1TS).
\end{itemize}
Here we define as S-star the single or binary star that remains bound to the MBH \citep{gou03,brw15}. We use energy consideration in the encounter in order to discrimate among the possible outcomes. We integrate the system for a total time $T=D/v$, where $v$ is the initial velocity of the centre of mass of the triple. This choice of the total integration time allows us to resolve all the possible channels for all the scattering events. At $t=T$, we first determine if the stars are in a triple system. If they are not, we determine if any pair of stars form a bound system. Once the hierarchy of the stars is known, we determine if each sub-system (single star, binary or triple) is bound or unbound to the MBH.

We initialise the initial conditions of the centre of mass of the binary following the prescriptions of \citet{gl06,gl07}. Each triple starts from a distance $D=10^3\times \aout$ with respect to the MBH. We fix the orbital plane of the centre of mass of the triple and set the initial transverse velocity to $v = 250\kms$ \citep{hills88}. \citet{yut03} show that the results of the scattering experiments are independent of the choice of initial velocity of the binary relative to the MBH as long as it is much smaller than the relative velocity at the minimum distance
\begin{equation}
v\lesssim 1.4\times 10^4\ \mathrm{km\ s}^{-1}\left(\frac{0.1\ \mathrm{AU}}{a}\right)\,.
\end{equation}
For $a=0.25 \au$ (the minimum semi-major axis of the outer star in the triple), the maximum allowed velocity is $\approx 2000\kms$, much larger than the typical dispersion velocity in the Galactic Centre. 
By using Eq.\,\ref{eqn:rmin}, we generate the maximum initial distance for which the pericentre of the triple is $\lesssim r_t$. We then randomly sample initial distances up to such maximum according to a probability $f(b)\propto b$ in the pericentre distance, as appropriate when gravitational focusing is important \citep{hills88,brm06}. The nature of the system is chaotic and depends on the relative inital phases of the inner binary, outer binary and triple orbit with respect to the MBH. Moreover, the relevant angles that define the triple's geometry are randomly sampled and the Kozai-Lidov oscillations in high-inclines systems could make the problem even more chaotic \citep{lid62,koz62}. All these ingredients make predictions of outcome probabilities based on simple analytical considerations unreliable.

\begin{table}
\caption{Models: name, star mass ($m_*$), star radius ($R_*$), initial inner binary semi-major axis ($\ain$), initial outer binary semi-major axis ($\aout$).}
\centering
\begin{tabular}{|l|c|c|c|c|}
\hline
Name & $m_*$ ($\msun$) & $R_*$ ($\rsun$) & $\ain$ (AU) & $\aout$ (AU)\\
\hline
Model 1		& $3$ 		& $0$	 & $0.05$-$0.1$		& $0.5$\\
Model 1r	& $3$ 		& yes	 & $0.05$-$0.1$ 	& $0.5$\\
Model 2		& $3$ 		& $0$ 	 & $0.05$			& $0.5$-$1.0$\\
Model 2r	& $3$ 		& yes	 & $0.05$			& $0.5$-$1.0$\\
Model 3		& $1$-$4$ 	& $0$	 & $0.05$			& $0.5$\\
Model 3r	& $1$-$4$ 	& yes	 & $0.05$			& $0.5$\\
Model 4		& $3$ 		& $0$	 & $0.025$-$0.05$	& $0.25$\\
Model 5		& $3$ 		& $0$ 	 & $0.025$			& $0.25$-$0.5$\\
Model 6		& $1$-$4$	& $0$	 & $0.025$			& $0.25$\\
\hline
\end{tabular}
\label{tab1}
\end{table}
The initial conditions for the numerical experiments have been set as follows  (see also Table\,\ref{tab1}):
\begin{itemize}
\item The mass of the MBH is fixed to $M=4\times 10^6\msun$ \citep{gil09}.
\item Stellar masses are set to $m_*=1$, $2$, $3$, $4 \msun$.
\item Stellar radii are computed from \citep{dem91}
\begin{equation}
R_*=
\begin{cases}
1.06\ (m_*/\msun)^{0.945}\rsun& \text{$ m_*< 1.66\msun$},\\
1.33\ (m_*/\msun)^{0.555}\rsun& \text{$ m_*> 1.66\msun$}.
\end{cases}
\end{equation}
All models marked with ``r'' have finite stellar radii taken into account. In this case, the relative distance of any two stars is monitored during the encounter. If any such distance becomes smaller than the sum of the stellar radii, the stars are considered merged and removed from the simulation.
\item The semi-major axis of the inner binary is $\ain = 0.025$-$0.1\au$.
\item The initial eccentricity of the inner and outer binaries is $e_{\rm in}=e_{\rm out}=0$.
\item The initial phase $\chi_{1}$ of the inner binary, which determines the initial position of the stars on the orbit, is randomly generated.
\item The angles $\theta_{1}$, $\phi_{1}$, $\psi_{1}$, which determine the orientation of the inner binary's orbital plane with respect to the  orbital plane of the centre of mass of the triple, are randomly generated.
\item The semi-major axis of the outer binary is $\aout = 0.25$-$1.0\au$. 
\item The initial phase $\chi_{2}$ of the outer binary, which determines the initial position of the outer star with respect to the inner binary, is randomly generated.
\item The angles $\theta_{2}$, $\phi_{2}$, $\psi_{2}$, which determine the orientation of the orbital plane of the outer star with respect to the orbital plane of the centre of mass of the triple, are randomly generated;
\item The initial distance of the triple from the MBH is $D=10^3\times \aout$.
\end{itemize}

Initial circular orbits are not a serious limitation (see also \citet{brm06}). In analogy to the binary tidal disruption, the outputs of the problem depend mainly on the energy reservoir of the triple (see Section 3). Hence, we argue that the results should be quite insensitive to the initial eccentricity of the inner and outer orbits. On the other hand, non-zero eccentricity would probably favour collisions between two stars of the triple since their relative distance at the orbital pericenter would be smaller then the circular case. We also note that our initial configuration satisfies the stability criterion of hierarchical triples \citep{mar01}
\begin{equation}
\frac{R_{p}}{\ain}\geq 2.8 \left[\left(1+\frac{m_3}{m_1+m_2}\right)\frac{1+e_{out}}{\sqrt{1-e_{out}}} \right]^{2/5}\ ,
\label{eqn:stabts}
\end{equation}
where $m_1$ and $m_2$ represent the masses of the inner binary stars, $m_3$ the mass of the outer star and $R_p$ its pericentre distance.
For our set up the criterion leads to $\aout/\ain\geq 3.3$, which is satisfied by our initial conditions. We run simulations with $\aout/\ain\ge 5$ to explore how this ratio affects the relative outcome probabilities.

Given the above set of initial parameters, we integrated the system of the differential equations of motion of the 4-bodies
\begin{equation}
{\ddot{\textbf{r}}}_i=-G\sum\limits_{j\ne i}\frac{m_j(\textbf{r}_i-\textbf{r}_j)}{\left|\textbf{r}_i-\textbf{r}_j\right|^3}\ ,
\end{equation}
with $i=1$,$2$,$3$,$4$, using the \textsc{ARCHAIN} code \citep{mik06,mik08}, a fully regolarised code able to model the evolution of binaries of arbitrary mass ratios with extreme accuracy, even over long periods of time. By combining a chain structure \citep{ma93} with a time transformation, the algorithm avoids singularities and produces extremely accurate trajectories. In our numerical experiments, the fractional energy error remains below $10^{-10}$ over the whole integration time.

\section{Results}

\begin{figure*} 
\centering
\begin{minipage}{20.5cm}
\subfloat{\includegraphics[scale=0.58]{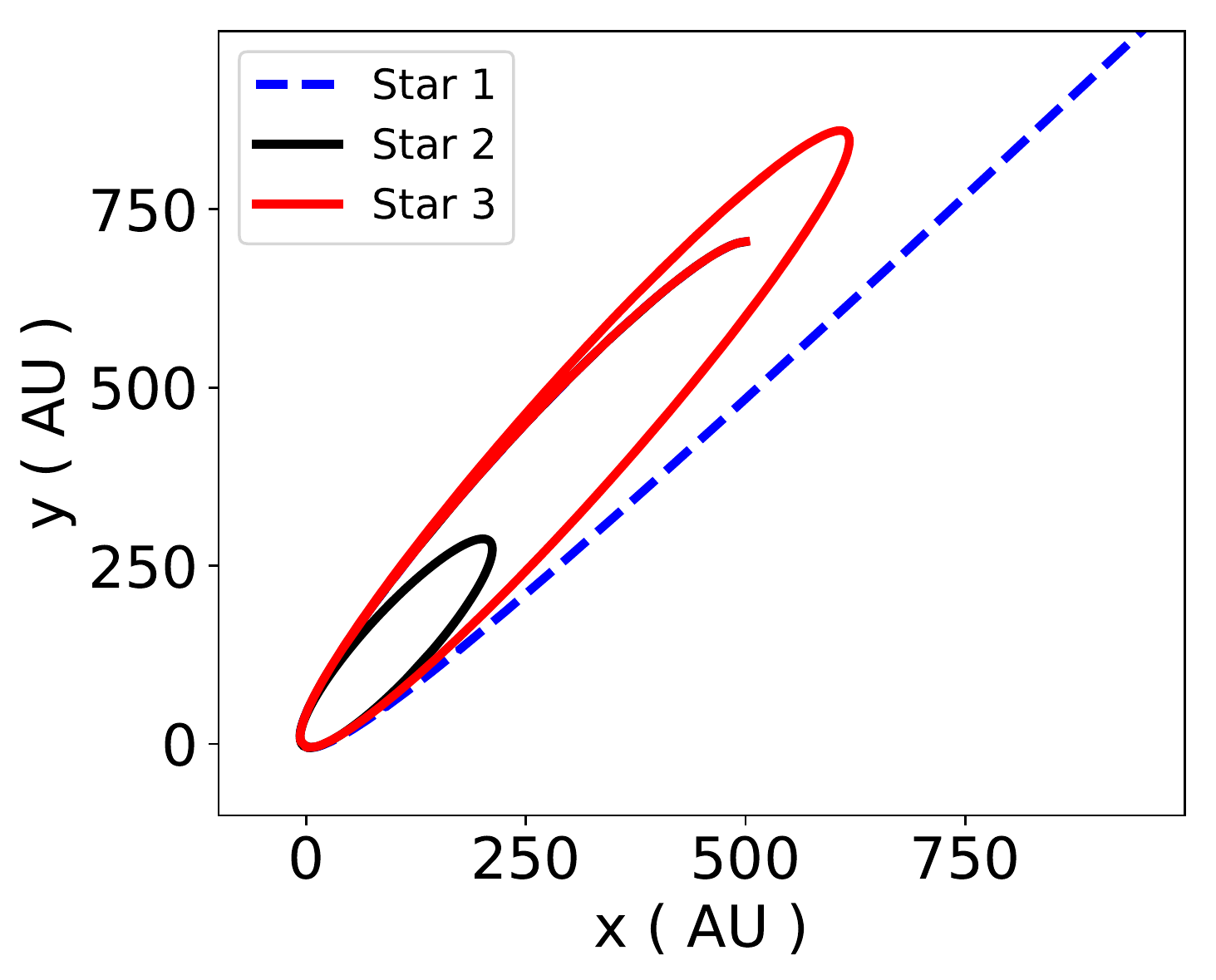}}
\subfloat{\includegraphics[scale=0.58]{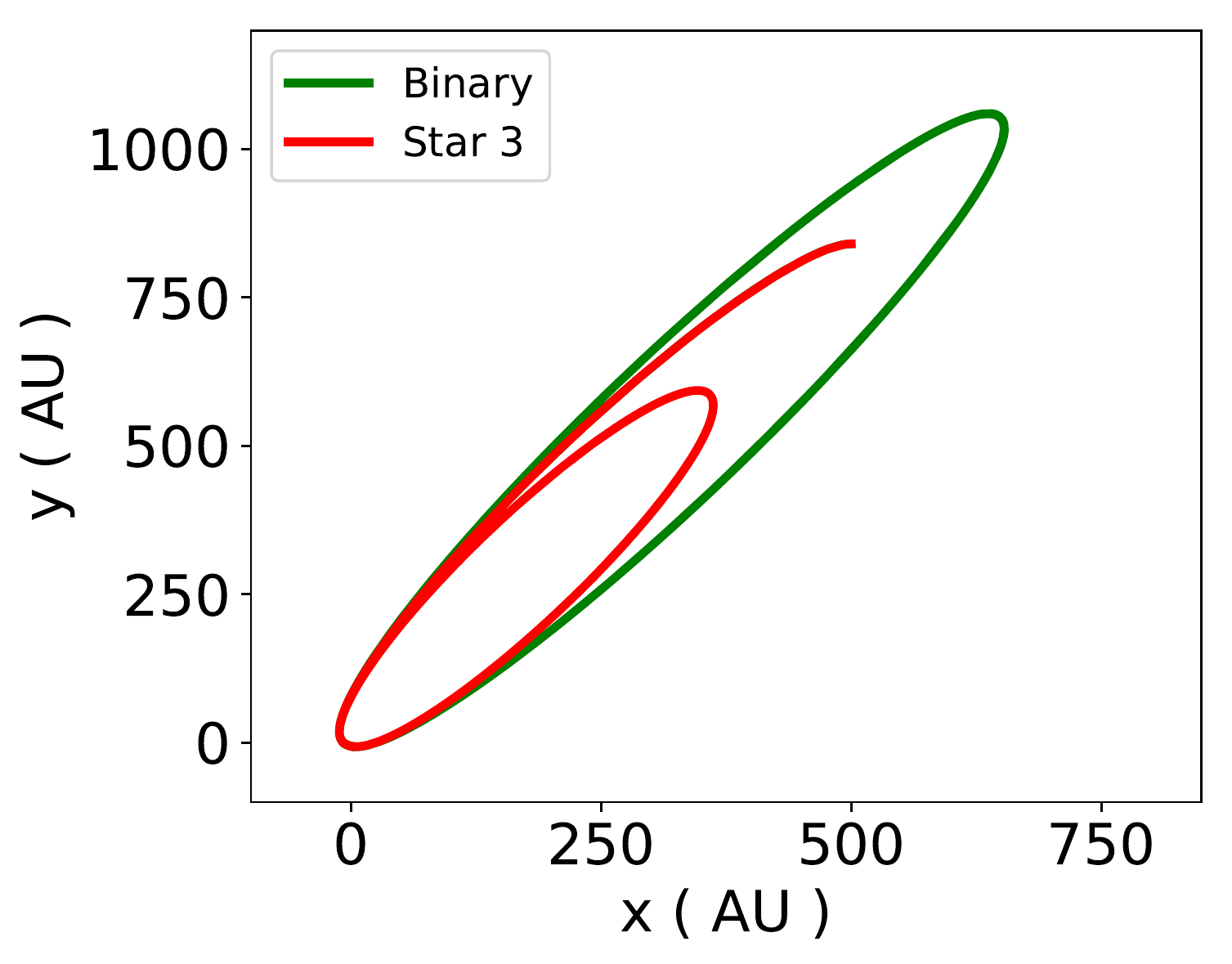}}
\end{minipage}
\caption{Examples of scattering for Model 1 in the case $\ain=0.05$ AU. The initial distance on the x-axis is $500$ AU, set by the initial outer binary semi-major axis $\aout=0.5$ AU, while the initial distance on the y-axis is set by the impact parameter. The MBH is at the origin of the reference frame. Left panel: the outer binary is disrupted leaving a single S-star, while the inner binary leads to the production of a single HVS and a second S-star. Right Panel: the outer binary is disrupted leaving a single S-star, while the inner binary remains bound and orbits the MBH on a bound orbit. 
Both of single and binary S-stars move on high eccentricity orbits.}
\label{fig:scattex}
\end{figure*}
We performed $10^4$ simulations of close encounters for each combination of the parameters given in Table\,\ref{tab1}, 
for a total of $4.2\times10^5$ experiments. 

In Model 1 (Model 1r) we study the fate of triples as a function of $\ain$, the inner binary semi-major axis, while fixing the outer binary semi-major axis to $\aout=0.5$ AU and masses to $m_*=3\msun$, with zero (finite) stellar radii. In Model 2 (Model 2r) we consider triples with different values of $\aout$ and fix $\ain=0.05$ AU and $m_*=3\msun$, with zero (finite) stellar radius. In Model 3 (Model 3r) triples have different initial masses $m_*$ with fixed semi-major axes $\ain=0.05$ AU and $\aout=0.5$ AU, with zero (finite) stellar radius. Finally, in Model 4/5/6 we consider the same initial conditions for triples as in models 1/2/3 but half the values for $\ain$ and $\aout$ (see Table\,\ref{tab1}).

Two examples of scatterings for Model 1 (with $\ain=0.09$ AU) are shown in Fig.\,\ref{fig:scattex}, resulting in one HVS and two S-stars (1SH-2SS, left panel) and one single and one binary S-star (1SS-2SS, right panel). All stars are left bound to the MBH with large eccentricity.

\begin{figure} 
\centering
\includegraphics[scale=0.6]{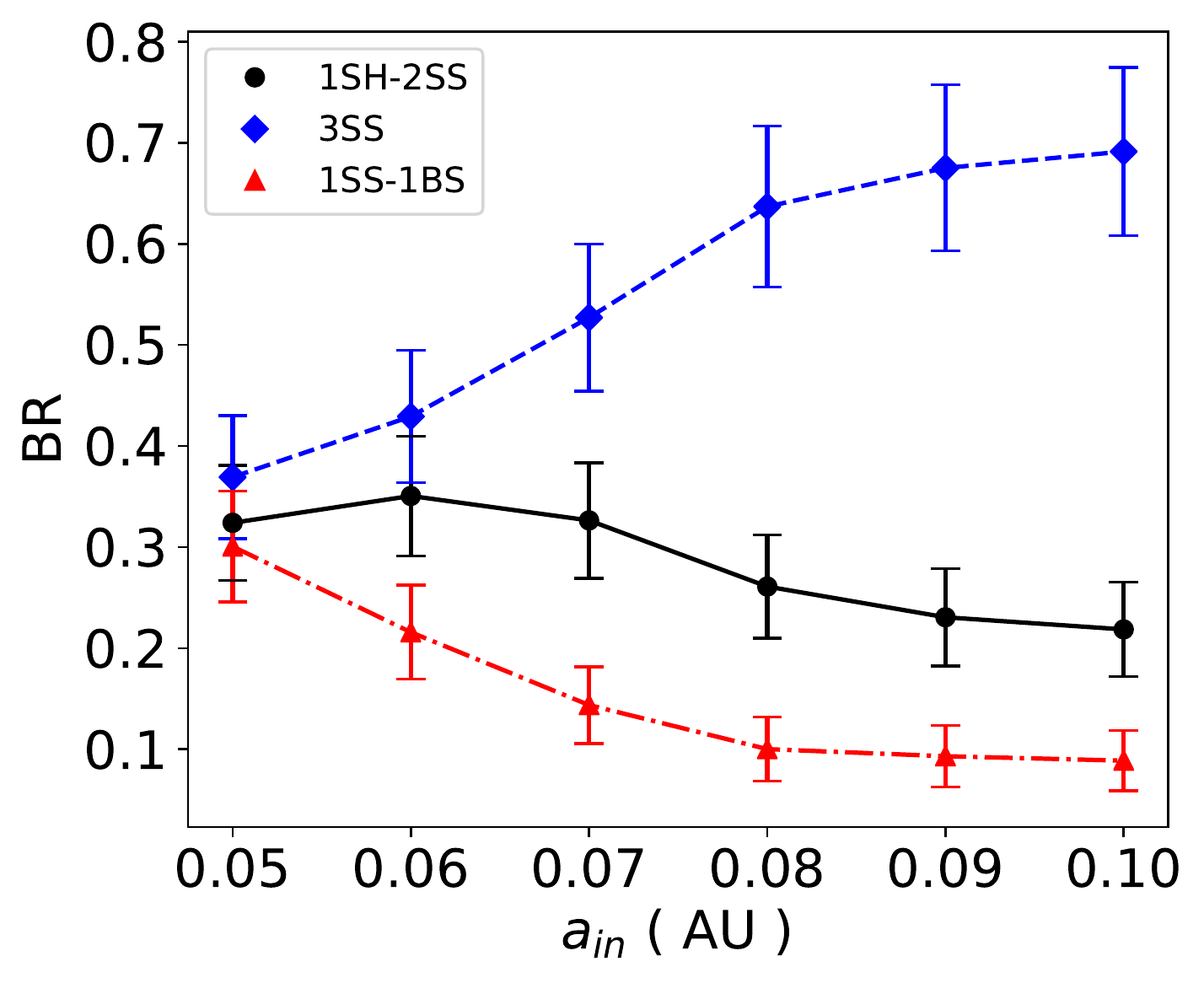}
\includegraphics[scale=0.6]{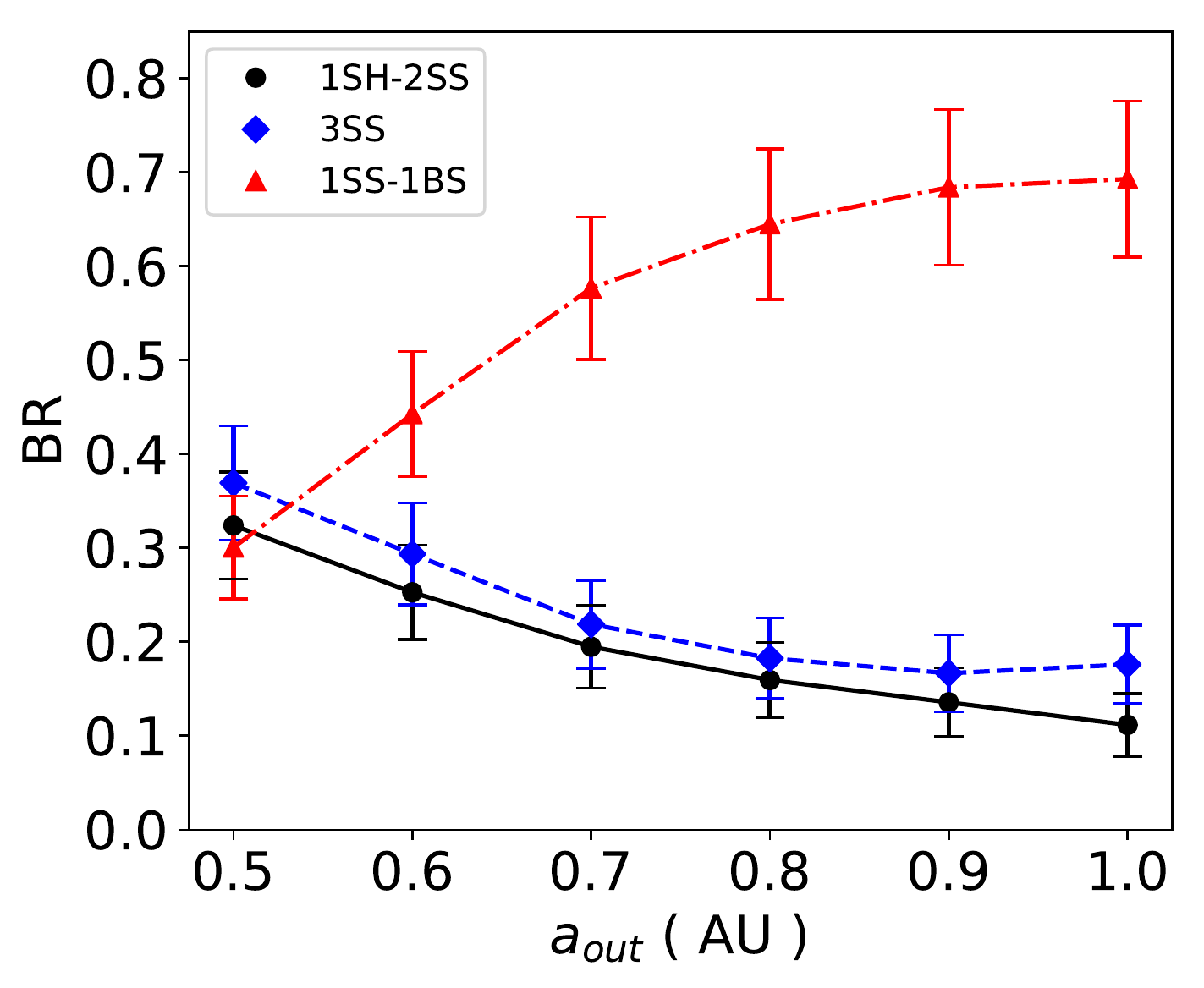}
\includegraphics[scale=0.6]{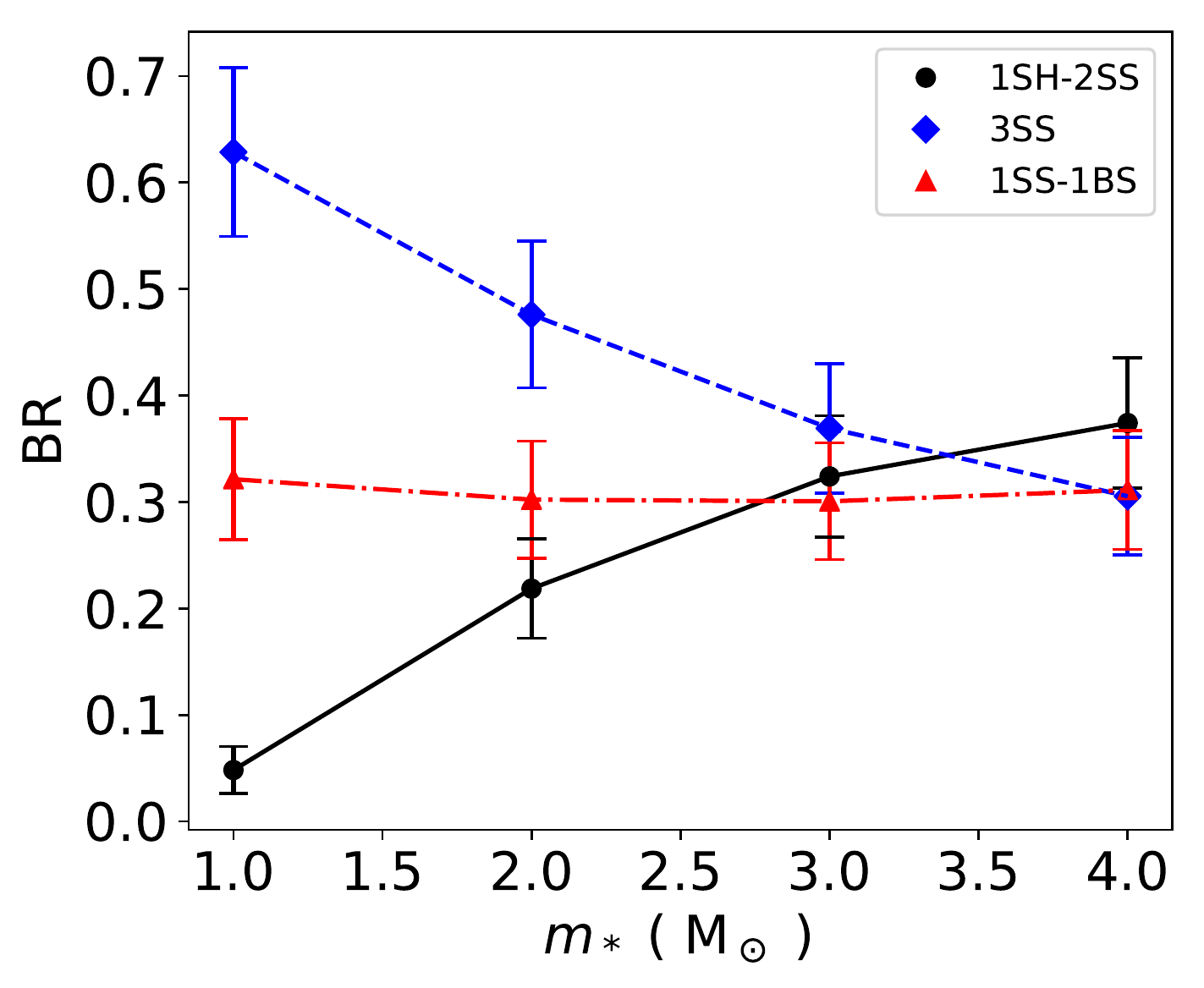}
\caption{Branching ratios for the different channels for Model 1 (top) as function of $\ain$, for Model 2 (centre) as function of $\aout$ and for Model 3 (bottom) as function of $m_*$. Poisson error bars are shown.}
\label{fig:br_123}
\end{figure}

\begin{figure} 
\centering
\includegraphics[scale=0.6]{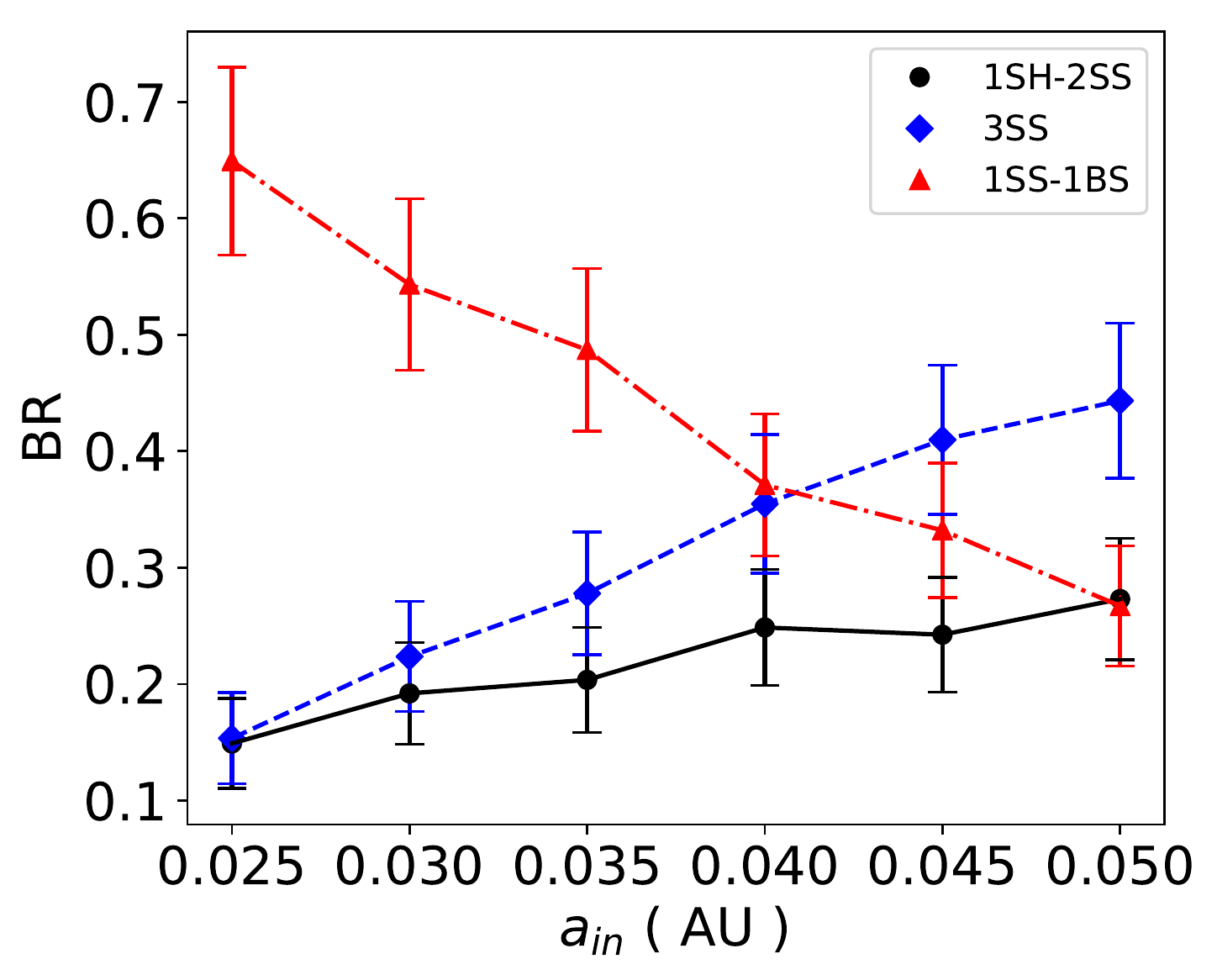}
\includegraphics[scale=0.6]{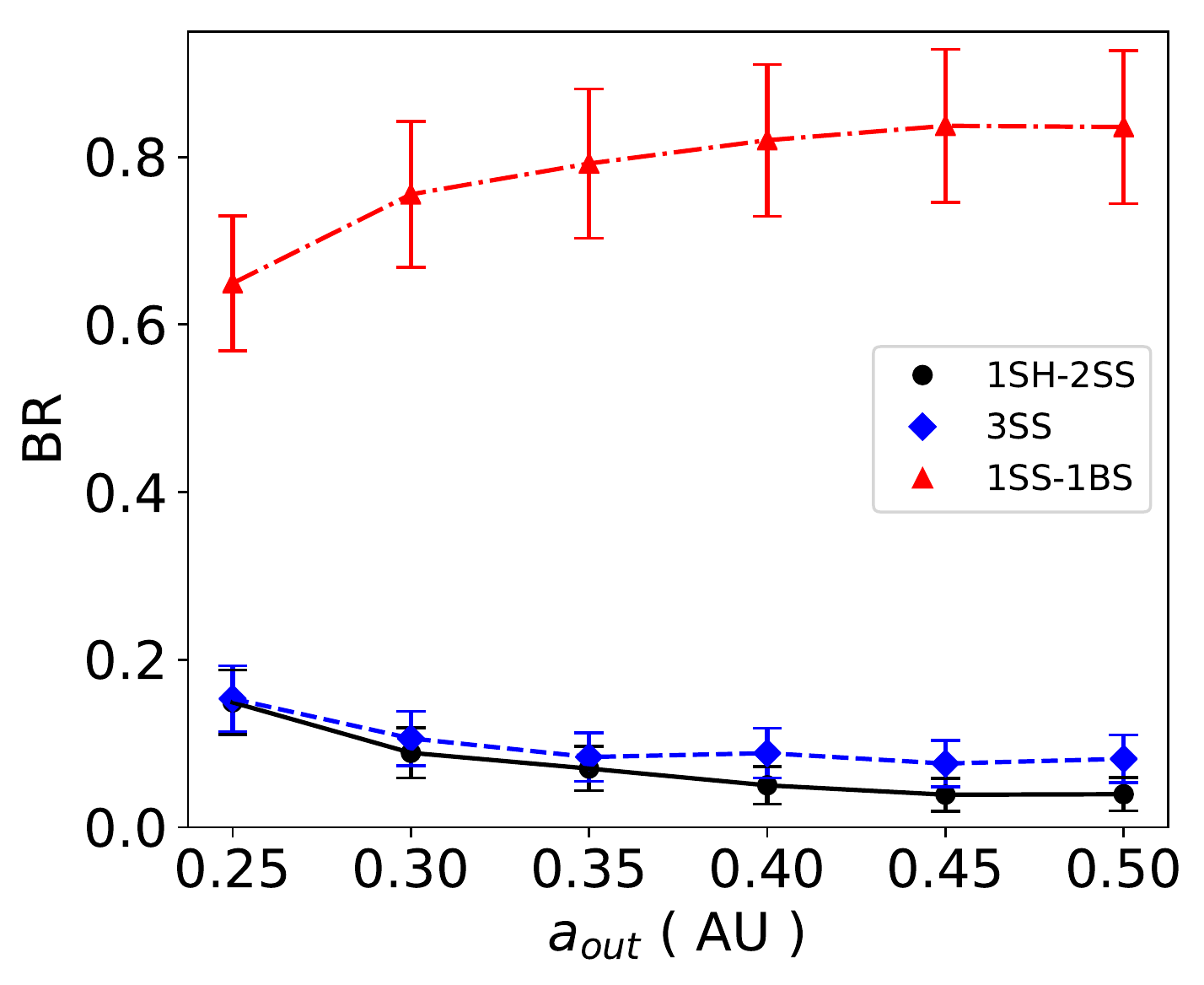}
\includegraphics[scale=0.6]{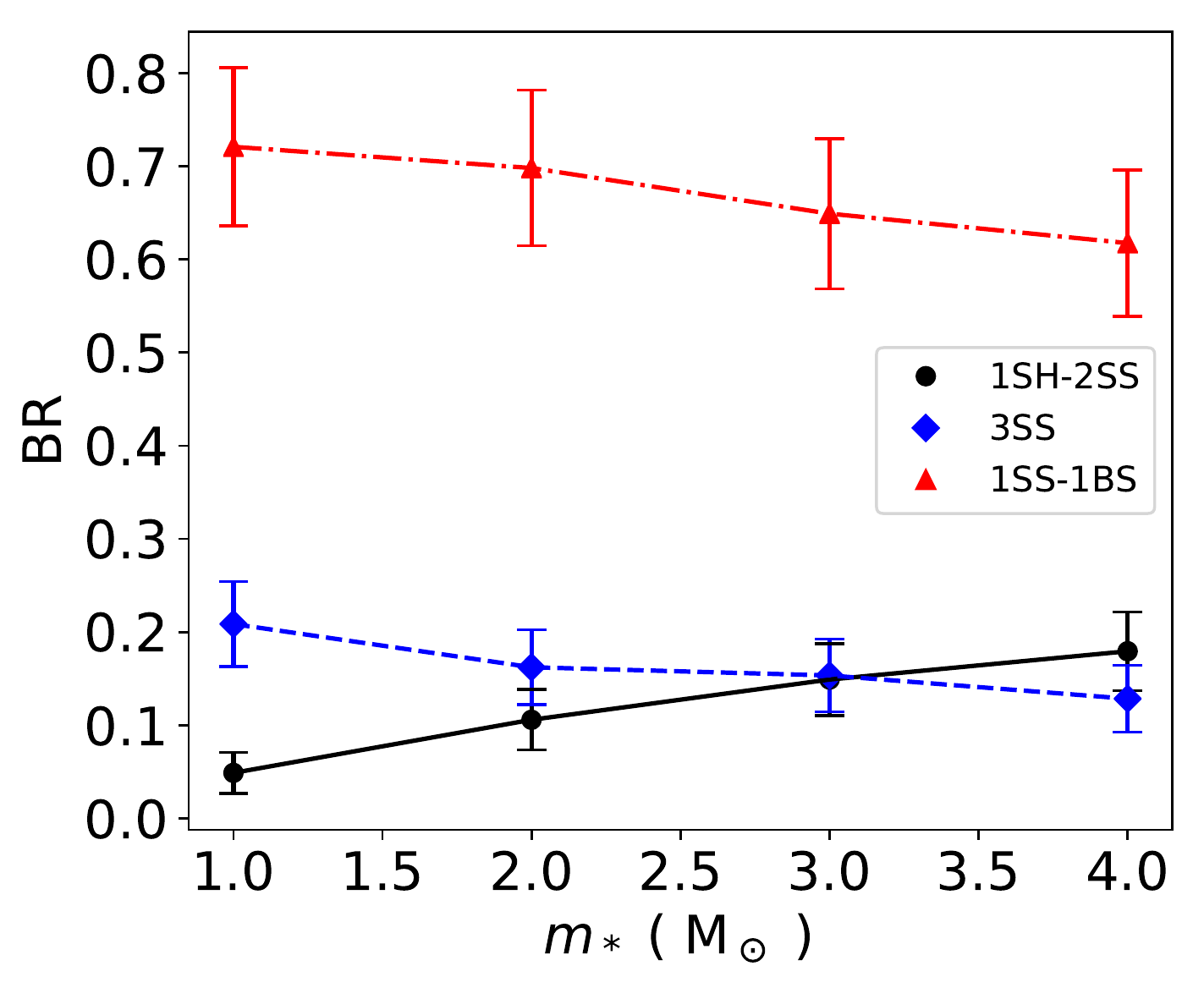}
\caption{Branching ratios for the different channels for Model 4 (top), Model 5 (central) and Model 6 (bottom) as function of $\ain$, $\aout$ and $m_*$, respectively. Poisson error bars are shown.}
\label{fig:br_456}
\end{figure}
The probabilities of different outcomes, the so-called ``branching ratios'' (BRs) are shown in Fig.\,\ref{fig:br_123} for models 1/2/3 (with the relative Poisson error bars), where we consider the dependence of the different outcomes as a function of $\ain$, $\aout$  and $m_*$. Only channels 1SH-2SS, 3SS and 1SS-1BS have significant probabilities and are shown in the figures, with all other outcomes having probabilities smaller than a few percent. 
\begin{table}
\caption{Branching ratios for Model 1 as function of $a_{ain}$.}
\centering
\begin{tabular}{|c|c|c|c|c|c|c|c|}
\hline
$\ain$ & 1SH-1BS & 2SH-1SS & 1BH-1SS & 1TS\\
\hline
$0.05$	& $1.5\times 10^{-2}$	        & $3\times 10^{-3}$	& $3\times 10^{-3}$	& $3.3\times 10^{-2}$\\
$0.06$	& $8\times 10^{-3}$		& $2\times 10^{-3}$	& $5\times 10^{-3}$	& $1.6\times 10^{-2}$\\
$0.07$	& $9\times 10^{-3}$		& $2\times 10^{-3}$	& $2\times 10^{-3}$	& $8\times 10^{-3}$\\
$0.08$	& $4\times 10^{-3}$		& $5\times 10^{-3}$	& $4\times 10^{-3}$	& $1\times 10^{-2}$\\
$0.09$	& $3\times 10^{-3}$		& $3\times 10^{-3}$	& $1\times 10^{-3}$	& $4\times 10^{-3}$\\
$0.1$	& $2\times 10^{-3}$		& $3\times 10^{-3}$	& $1\times 10^{-3}$	& $4\times 10^{-3}$\\
\hline
\end{tabular}
\label{tab:br1}
\end{table}
Table\,\ref{tab:br1} reports the BRs for Model 1 as function of $\ain$. 
In Model 1, the probabilities for outcomes 1SH-2SS and 1SS-1BS are decreasing functions of $\ain$, while the probability for channel 3SS increases for larger inner semi-major axes. 
In Model 2, the BRs for channel 1SH-2SS and channel 3SS slightly decrease with larger outer semi-major axes, while the outcomes 1SS-1BS 
becomes more likely for larger $\aout$. 
Finally, in Model 3, the production of 1SS-1BS is nearly constant with stellar mass, while the probability for 3SS decreases and that of 
1SH-2SS increases with larger $m_*$. 
The probability of producing HVBs is very small, $\lesssim 1\%$ in all cases (see also Tab.\,\ref{tab:br1}). 

We find similar trends for Model 4/5/6, whose BRs are shown in Fig.\,\ref{fig:br_456} (with the relative Poisson error bars). In these models, the values of  $\ain$ and $\aout$ are half the values used in Model 1/2/3 and the triples are set up with $\aout/\ain\ge 5$ to ensure their stability (see Eq.\,\ref{eqn:stabts}).
All channels not shown in the figure have probabilities $\lesssim 1\%$, included the production of HVBs. 

Mergers occur in about $10-35\%$ of the encounters if finite stellar radii are taken into account. For example, collisions have a  probability of $\approx 35\%$ in Model 1 with $\ain=0.05$ AU, and the relative BRs of the different channels, included HVB production, are reduced with respect to the point mass cases shown in Fig.\,\ref{fig:br_123} and Tab.\,\ref{tab:br1}. We find that the collision probability decreases as the inner binary becomes wider, as expected.

We can interpret the previous results by means of the typical energy variations involved in the triple disruption scenario. The process of breaking up a triple has two well defined scales. The first scale is set by the tidal radius of the outer binary
\begin{equation}
r_{t,out}\approx \left(\frac{M}{m}\right)^{1/3} \aout\ .
\end{equation}
The second one is given by the tidal radius of the innermost star
\begin{equation}
r_{t,in}\approx \left(\frac{M}{m}\right)^{1/3} \ain=\beta r_{t,out}\ ,
\end{equation}
where $\beta=\ain/\aout$. A triple undergoing an encounter with the MBH with a pericentre distance $b$ in the range $[r_{t,in}-r_{t,out}]$
should be broken up into inner binary plus outer star, i.e. the outermost star in the system should be unbound from the innermost pair.

We can count the number of objects having $r_{t,in}<b<r_{t,out}$ according to the pericentre distance distribution $f(b)\propto b$ \citep{hills88,brm06}:
\begin{equation}
N(r_{t,in}<b<r_{t,out})=\int_{r_{t,in}}^{r_{t,out}} f(b)\ db=1-\beta^2\ .
\end{equation}
To satisfy Eq.\,\ref{eqn:stabts}, it must be $\beta^{-1}\ge 5$. Hence, $N(r_{t,in}<b<r_{t,out})\gtrsim 95\%$ and only $\lesssim 5\%$ of triples have pericentre distance $\le r_{t,in}$. As a consequence, a typical scattering would lead to the disruption of the triple by removing the outer object. When this happens, the increase in specific energy of the binary is of the order \citep{yut03}
\begin{equation}
\delta E\approx v \delta v\approx \left[\left(\frac{GM_{BH}}{b}\right)\left(\frac{Gm}{\aout}\right) \right]^{1/2}\ .
\end{equation}
This extra energy is converted into internal energy of the binary itself, which becomes wider. The previous equation can be rewritten in terms of the specific binding energy of the binary $E_b=-Gm/2 \ain$ as
\begin{equation}
\delta E\approx 100\alpha^{1/2} \frac{Gm}{\ain}=100\alpha^{1/2} E_b\ .
\end{equation}
Here we have assumed that $b=\alpha r_{t,out}$, where $\beta\le \alpha \le 1$. In our simulations, $5\le \beta^{-1}\le 10$ and $\delta E\gtrsim 32 E_b\gg E_b$. We can estimate the minimum value of $\beta$ by calculating the minimum $\ain$ and the maximum $\aout$. The former is set by the finite size of the stars. For solar mass stars, we have $\ain \approx 0.01$ AU, while $\ain \approx 0.02$ AU in the case $m_*=3\msun$. To estimate the maximum $\aout$, we can introduce the dimensionless parameter
\begin{equation}
\zeta=\frac{|E_b|}{\sigma^2}\ ,
\label{eqn:mamax}
\end{equation}
where $\sigma$ is the stellar velocity dispersion. Binaries with $\zeta\ll 1$ (soft binaries) are on average disrupted by the background population as a consequence of scattering events, while binaries with $\epsilon\gg 1$ (hard binaries) typically tend to become harder and can be dissolved by the MBH \citep{hop09}. Computing $\zeta=1$ at the MBH influence radius ($r_h\approx 2$ pc in the Milky Way) provides an indication of the critical separation $\hat{a}$ that marks the transition between the two regimes. In the case of our Galaxy, $\hat{a}=0.1(m_*/\msun)$ AU \citep{frasar17}. Assuming that all the triples with $\aout\ge 5\hat{a}$ are disrupted, $\delta E\approx 14 E_b$ and $\approx 11 E_b$ for $m_*=1\msun$ and $m_*=3\msun$, respectively.

Let us consider what happens during a typical scattering encounter. In most cases, the triple's impact parameter is such that the outer binary is disrupted. The lighter component may have a larger probability to be captured by the MBH \citep{brm06}, while the heavier to be ejected. As a consequence, the outer star typically becomes an S-star. The inner binary tends to have an increase $\delta E$ in its specific energy, which can be distributed among the different degrees of freedom of the binary itself. 
Since this shift is several times larger than $E_b$, the binary's semi-major axis can change significantly even if a small fraction of $\delta E$ is converted into internal energy. As a consequence, the binary widens and can be more easily disrupted by the tidal field of the MBH. Even if the inner binary is tidally broken up by the MBH, the process does not necessarily lead to the production of an HVS since the HVS ejection velocity $v_{ej}\propto a^{-1/2}$ \citep{hills88,brm06}. 

If enough energy is converted into binary internal energy to allow for tidal disruption, the inner binary is broken up and two bound S-stars are left. If, on the other hand, the change in $\ain$ is not sufficient for tidal disruption, the inner binary can either remain on a bound orbit around the MBH or be ejected as an HVB. We expect the latter channel to be disfavoured with respect to the former since $v_{ej}\propto (0.1\ \mathrm{AU}/\aout)^{-1/2}$ \citep{brm06}. However, we note that the system is by definition chaotic and particular combinations of relative initial phases and orbital inclinations 
may favour the ejection of HVBs. According to this simple model, only the channels 1SH-2SS, 3SS and 1SS-1BS have large BRs, as we find in our simulations. A more detailed analysis of the fate of the single HVSs and S-stars seems to support this finding. For example, in Model 1 with 
$\ain=0.05$ AU, we find that only $4\%$ of the single HVSs were the outermost stars in the original triple and that $98\%$ of the S-type binary stars are made up of the original inner binary. We note that the model holds for other masses smaller than about $10^8\msun$, above which stars are swallowed by the MBH instead of being tidally disrupted. So there are no HVSs above this MBH mass. For smaller masses, the model should hold but due to the chaotic nature of the processes new numerical experiments should be performed. We argue that also in the case of other galaxies, as the LMC, the three largest Branching Ratios would be 1SH-2SS, 3SS and 1SS-1BS, but their relative magnitude would differ from the Milky Way case.

\begin{figure*} 
\centering
\begin{minipage}{20.5cm}
\subfloat{\includegraphics[scale=0.6]{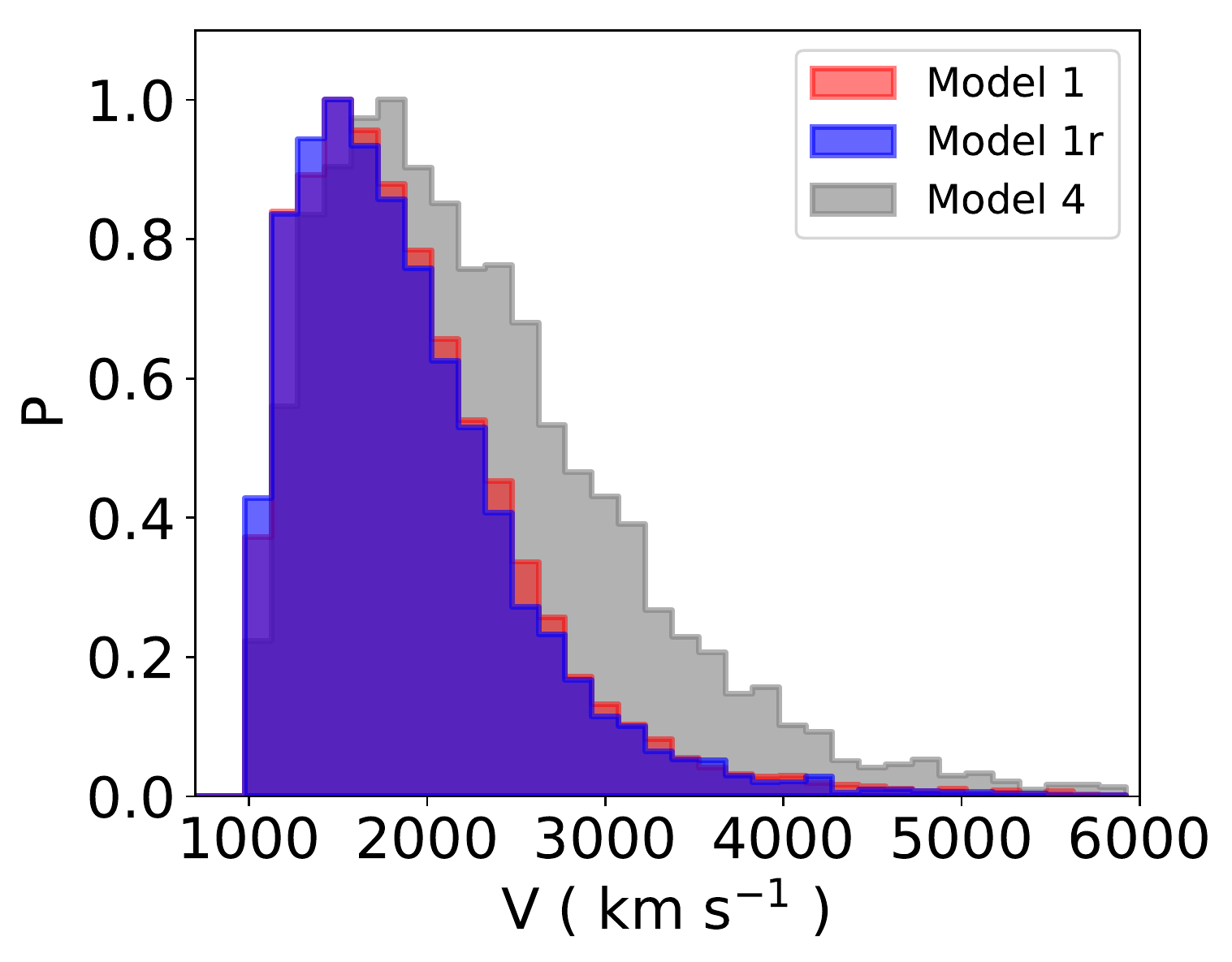}}
\subfloat{\includegraphics[scale=0.6]{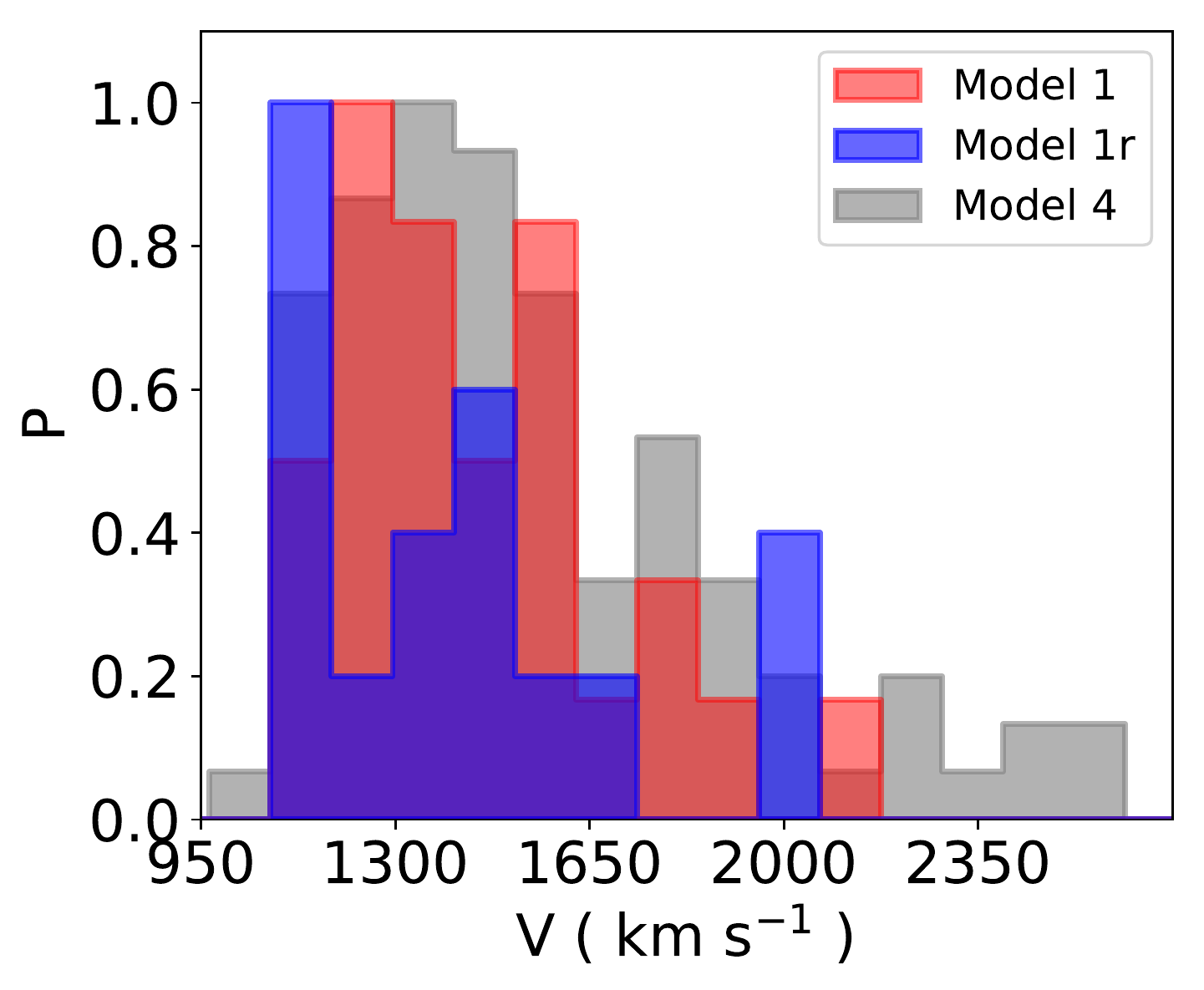}}
\end{minipage}
\begin{minipage}{20.5cm}
\subfloat{\includegraphics[scale=0.6]{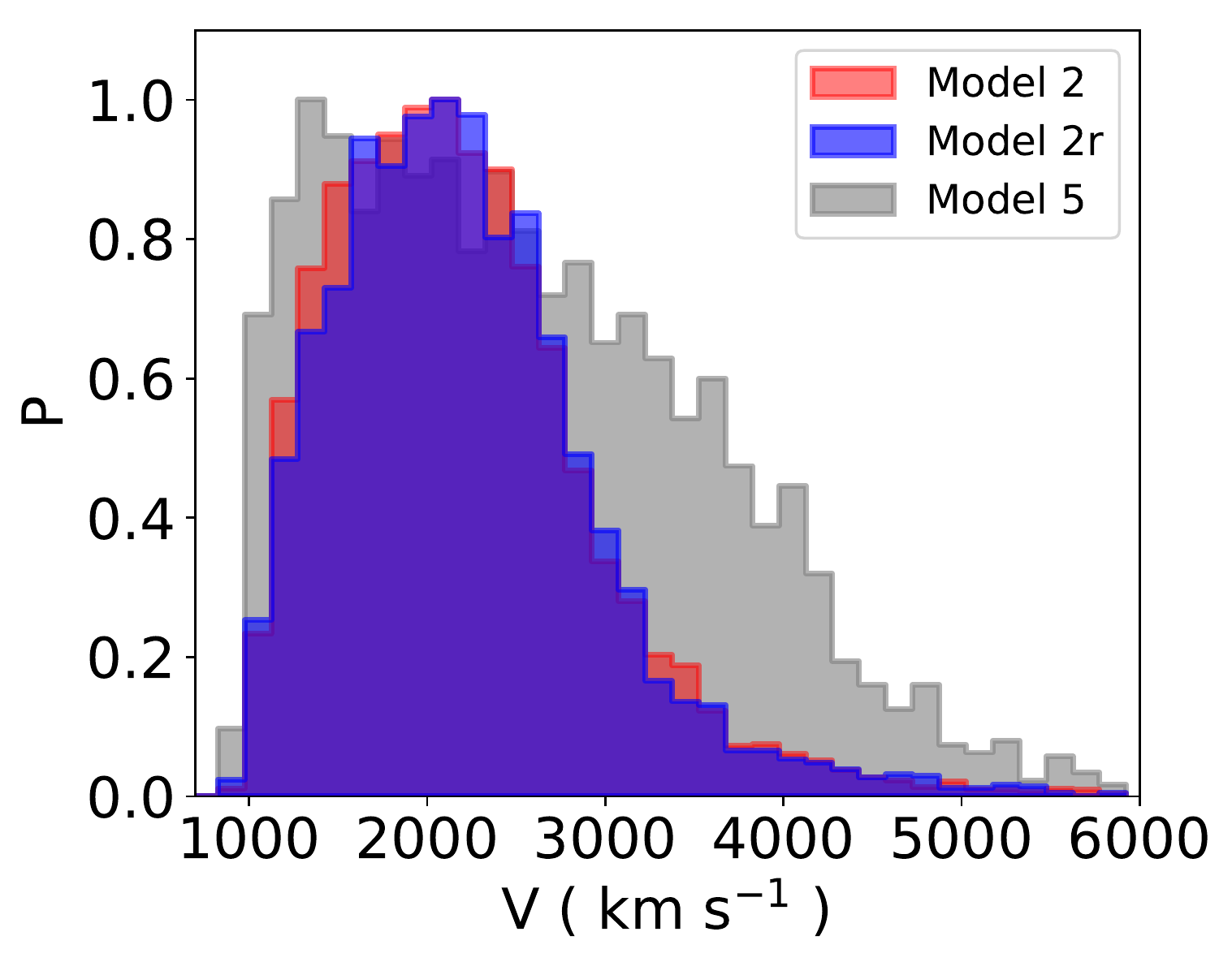}}
\subfloat{\includegraphics[scale=0.6]{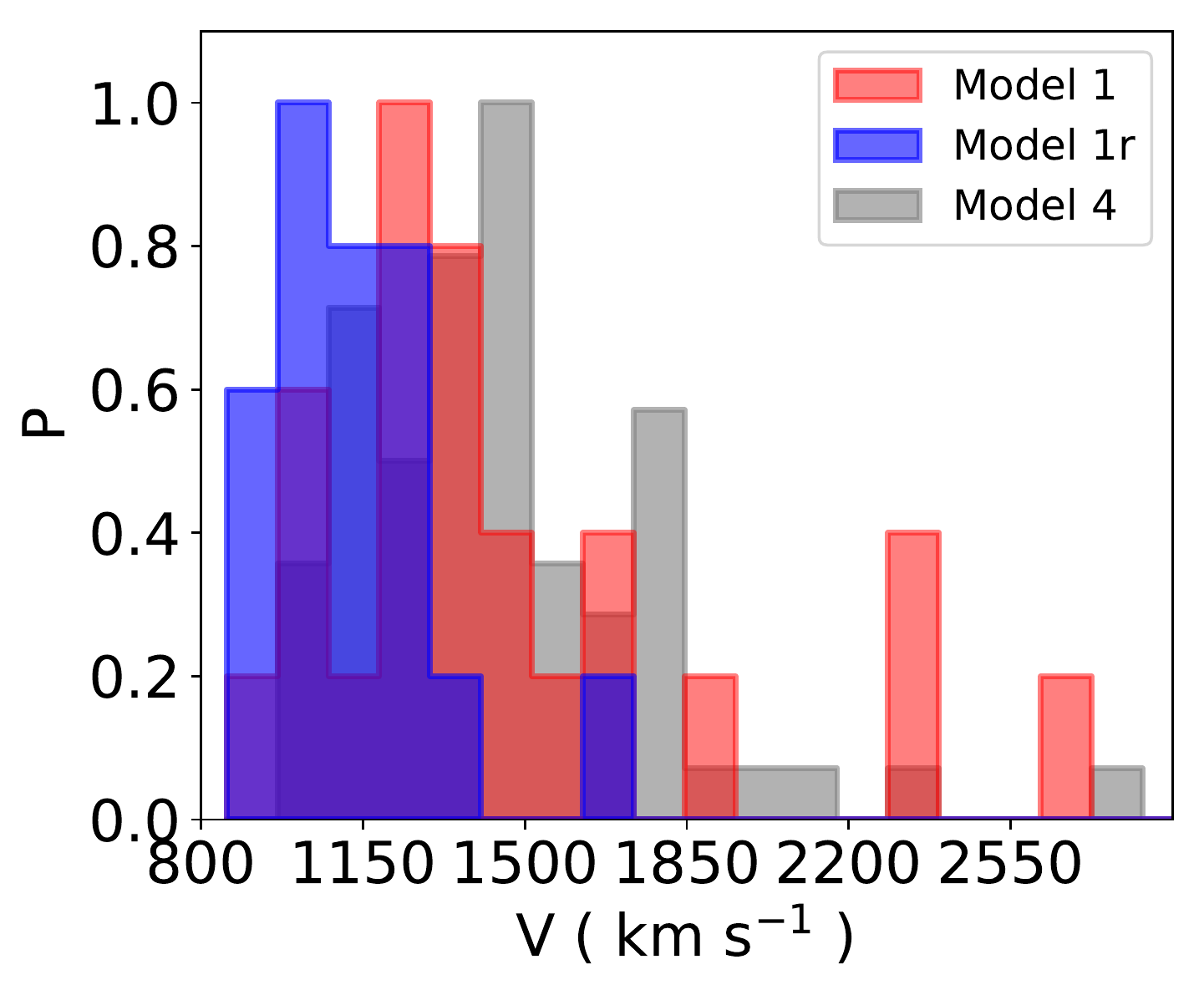}}
\end{minipage}
\begin{minipage}{20.5cm}
\subfloat{\includegraphics[scale=0.6]{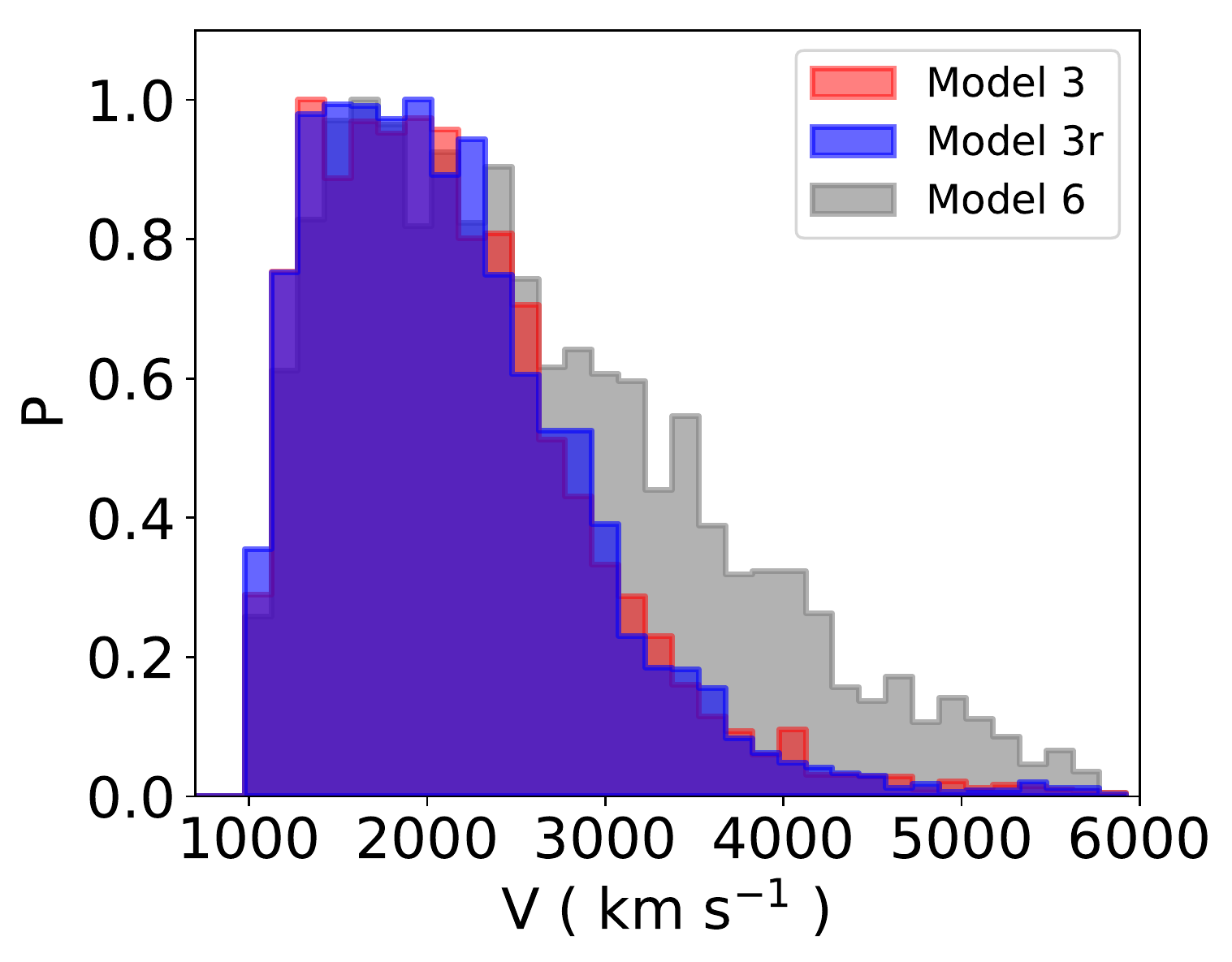}}
\subfloat{\includegraphics[scale=0.6]{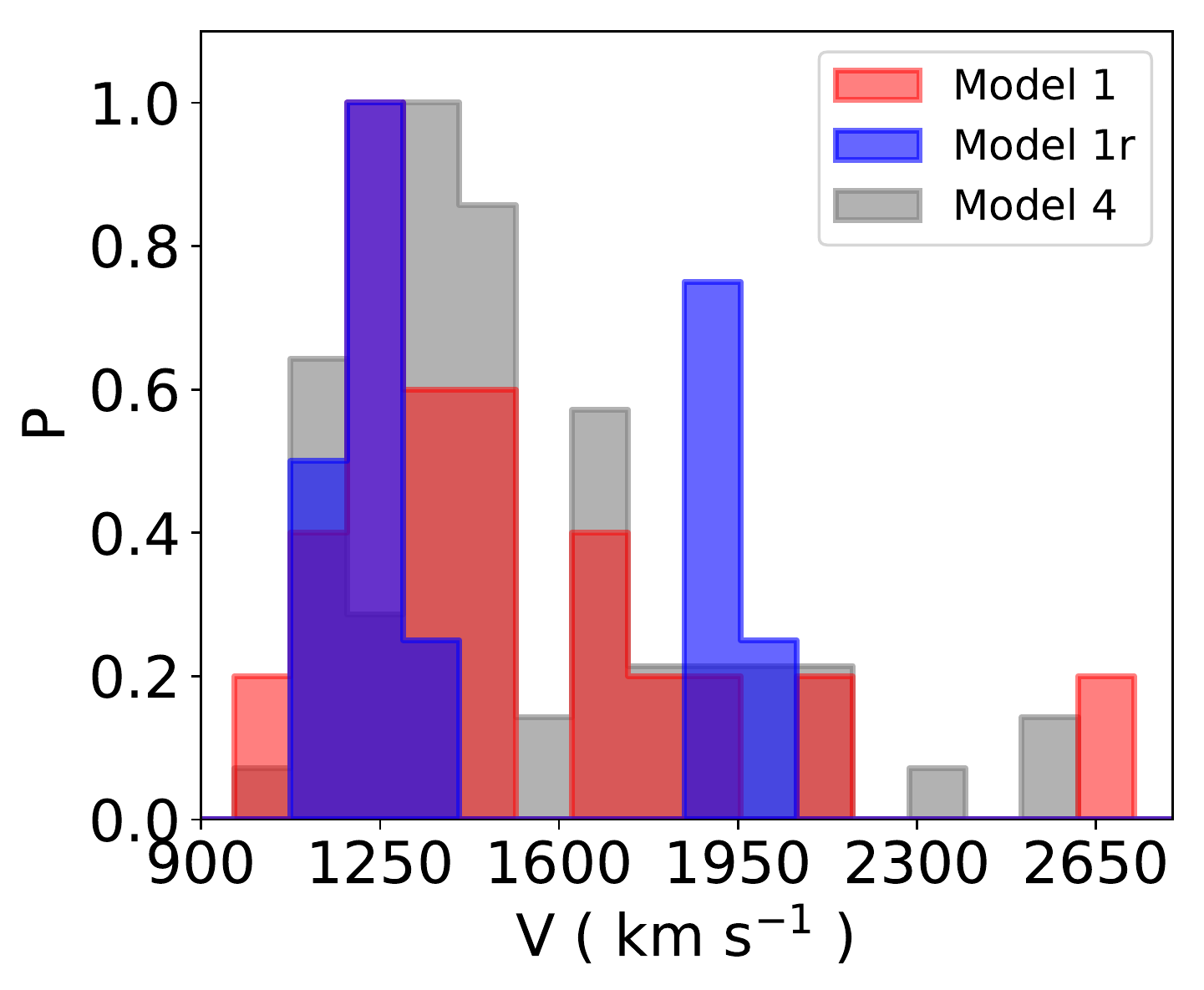}}
\end{minipage}
\caption{Velocity distribution for HVSs (left) and HVBs (right) for Model 1/1r/4, Model 2/2r/5 and Model 3/3r/6 when all the $\ain$, $\aout$ and $m_*$ are considered, respectively. The linestyles are the same on the left-hand side panels and right-hand side panels.}
\label{fig:velocity}
\end{figure*}

In Fig.\,\ref{fig:velocity} we show the velocity distribution for HVSs and HVBs in the different models, when all simulations with different $\ain$, $\aout$ and $m_*$ are considered. 
The distribution for HVSs is peaked around $2000\kms$ in all models, with a tail extending up to $6000\kms$. For Models 4-5-6, the distribution has a larger fraction of stars with velocities $\gtrsim 2000\kms$ as a consequence of the smaller initial $\ain$ and $\aout$.
The velocities of the few HVBs produced in the simulations are concentrated near the peak of $\approx 1300\kms$ with outliers up to $\approx 2600\kms$.

\begin{figure*}
\includegraphics[width=5.8cm]{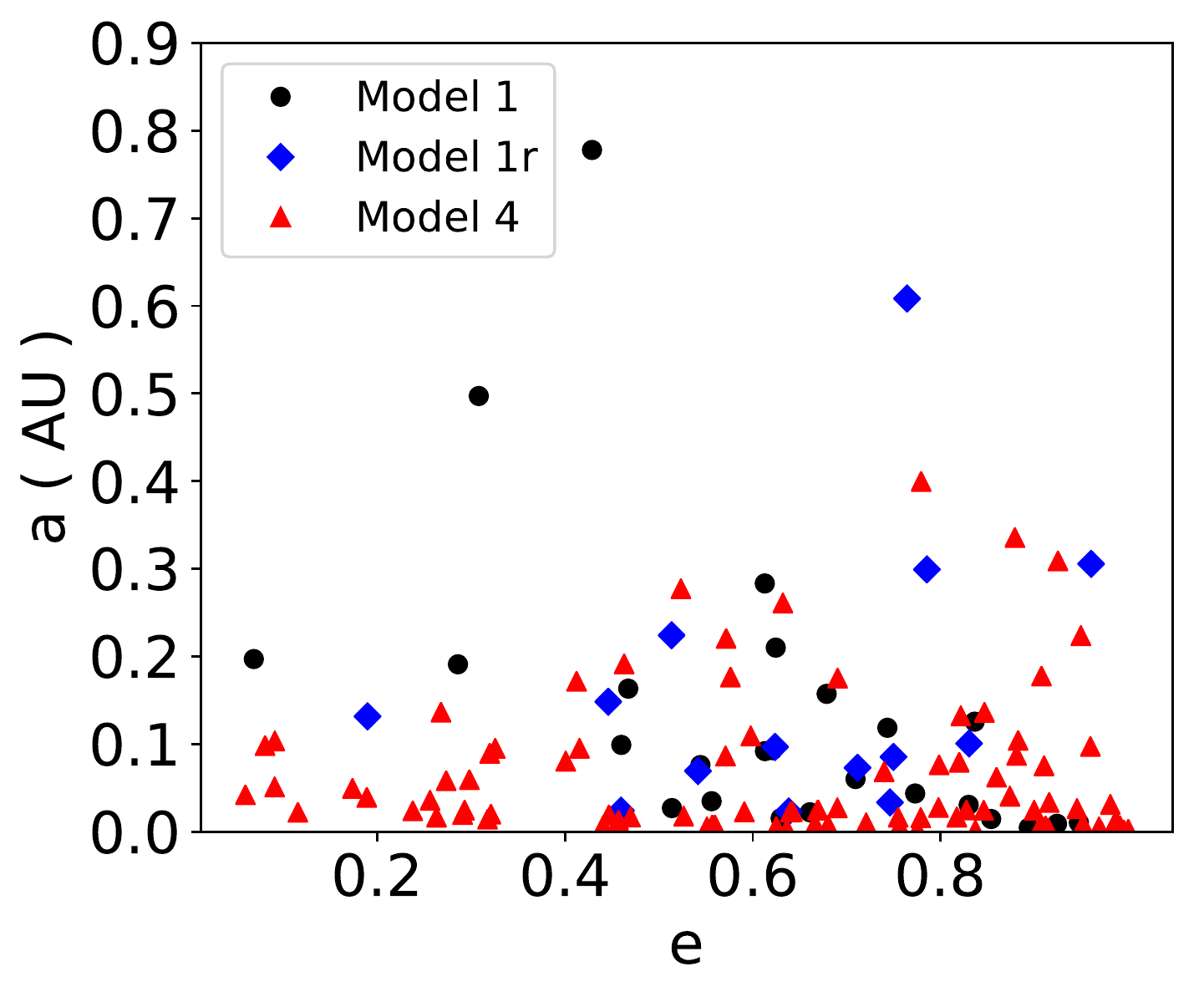}
\includegraphics[width=5.8cm]{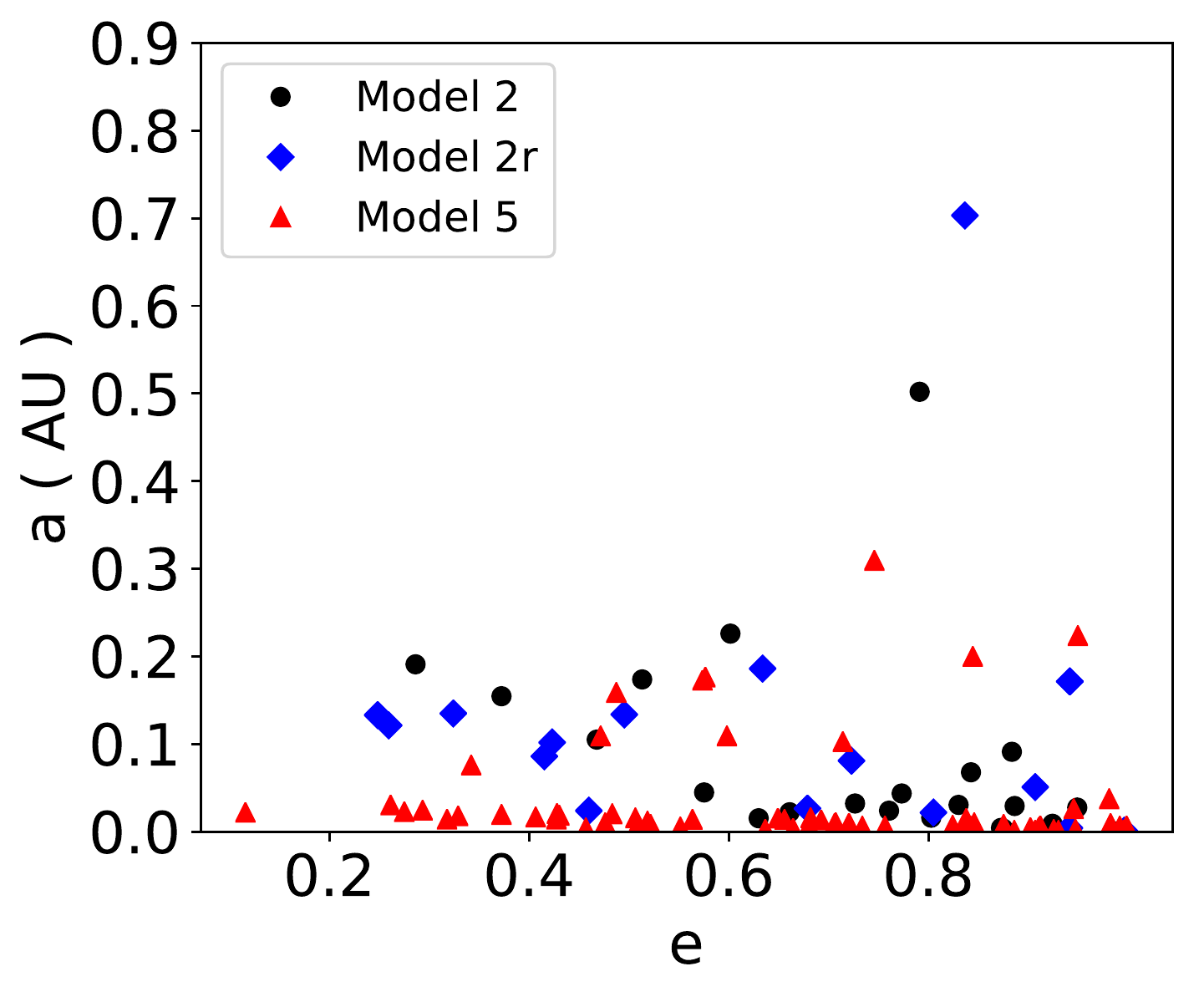}
\includegraphics[width=5.8cm]{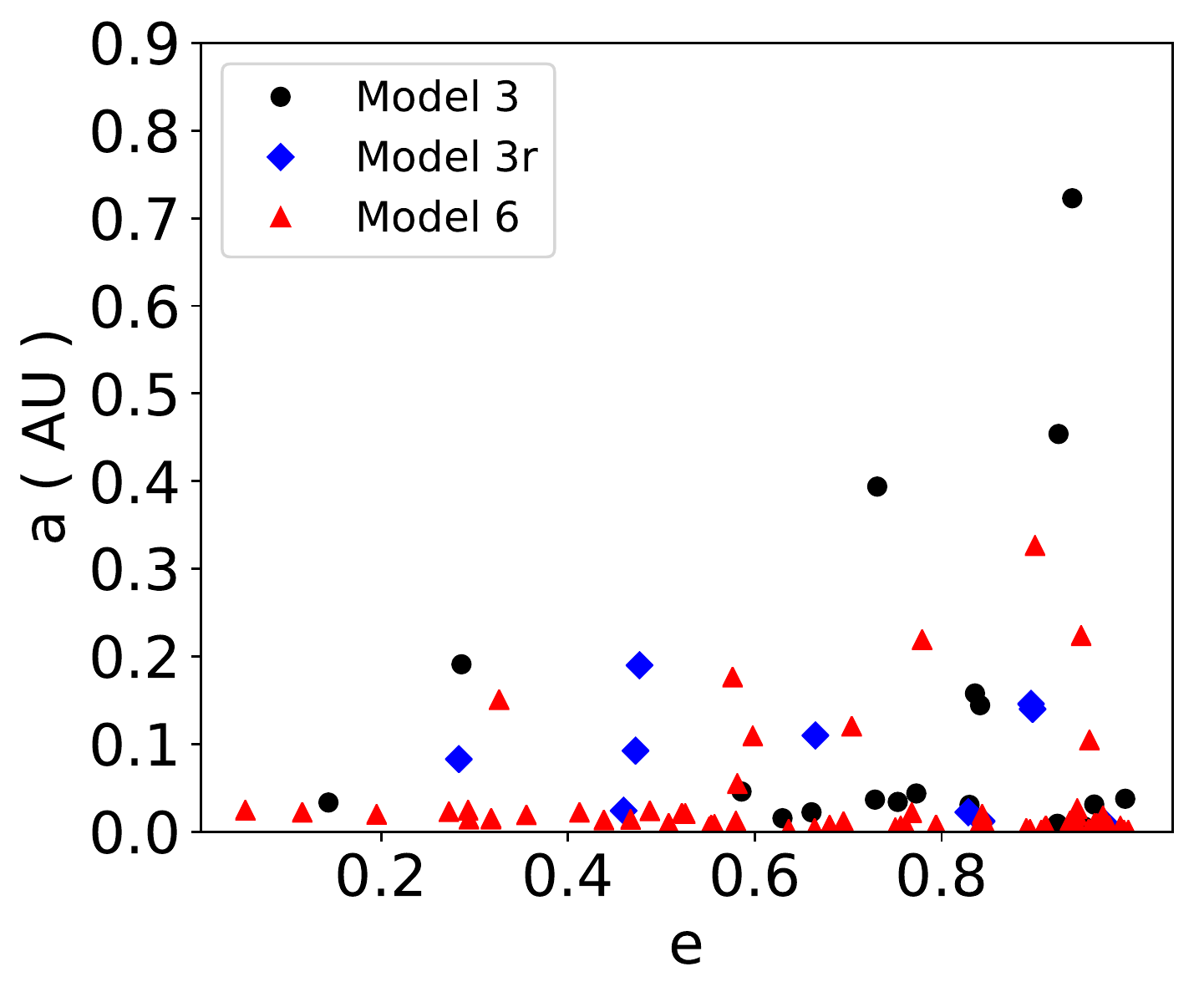}
\caption{Orbital parameters (semi-major axis and eccentricity) for all single HVBs produced in Model 1/1r/4 (left), Model 2/2r/5 (centre) and Model 3/3r/6 (right).}
\label{fig:binaryae}
\end{figure*}
Figure \ref{fig:binaryae} shows the semi-major axes and eccentricities of the HVBs produced as a consequence of triple disruption for all the models considered. It is clear that most of the HVBs have small semi-major axis ($\lesssim 0.2$ AU) and large eccentricity ($\gtrsim 0.5$). 
Based on the previous theoretical considerations, we expect most of them to be made up of the original inner binary, with a semi-major axis somewhat larger than the initial one. This is consistent with our experiments, where only $\approx 13\%$ and $\approx 8\%$ of the HVBs are made up of an exchanged binary (i.e. one composed of a star from the inner binary and the outermost star) in Models 1/2/3 and Models 4/5/6, respectively. We note that Models 4/5/6 produce about three times more HVBs than Models 1/2/3 as a consequence of the smaller initial inner and outer semi-major axis. The tighter the inner binary, the larger the energy resevoir that can be exchanged during the four-body encounter and the larger the production probability of HVBs.

\begin{figure*} 
\centering
\begin{minipage}{20.5cm}
\subfloat{\includegraphics[scale=0.6]{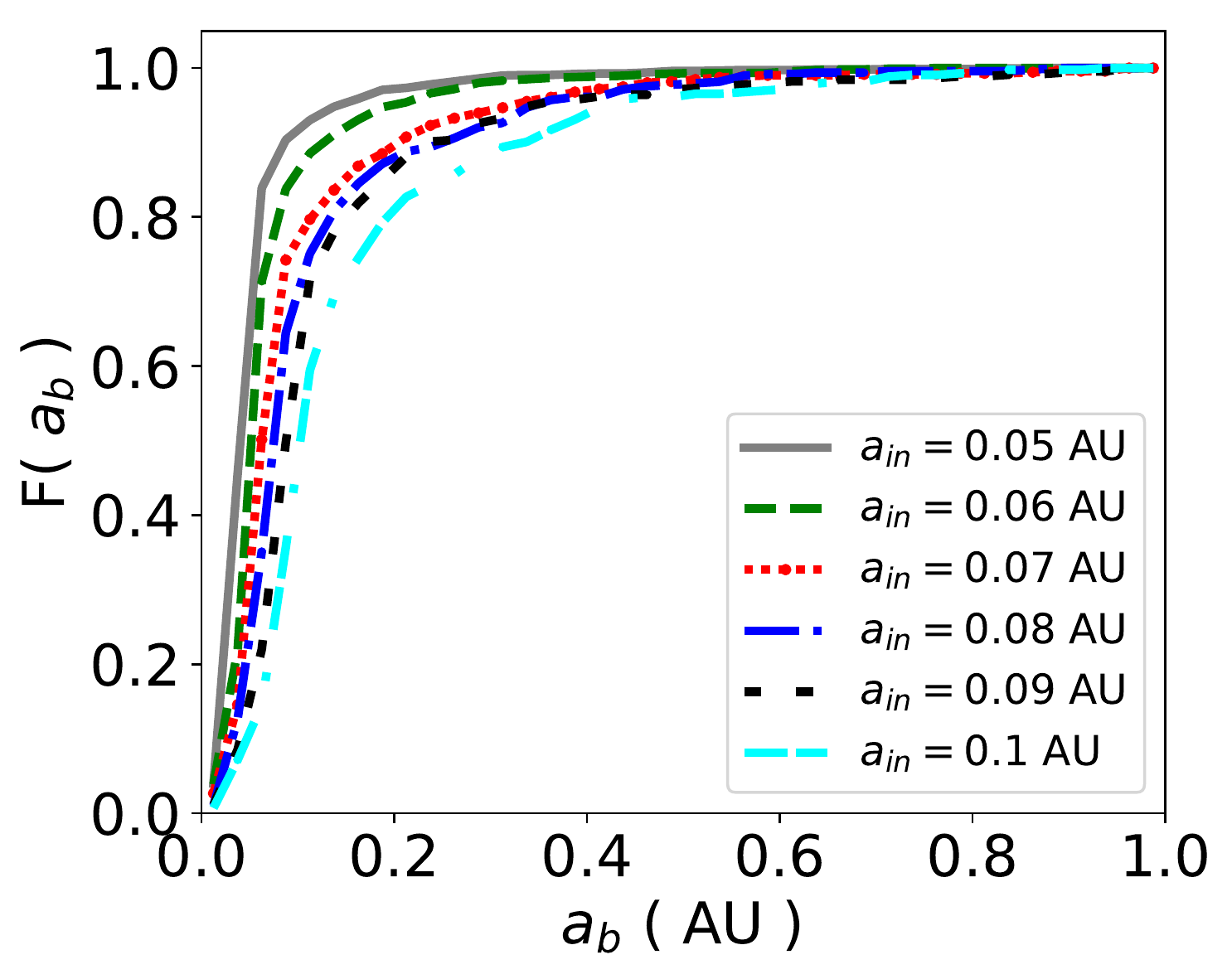}}
\subfloat{\includegraphics[scale=0.6]{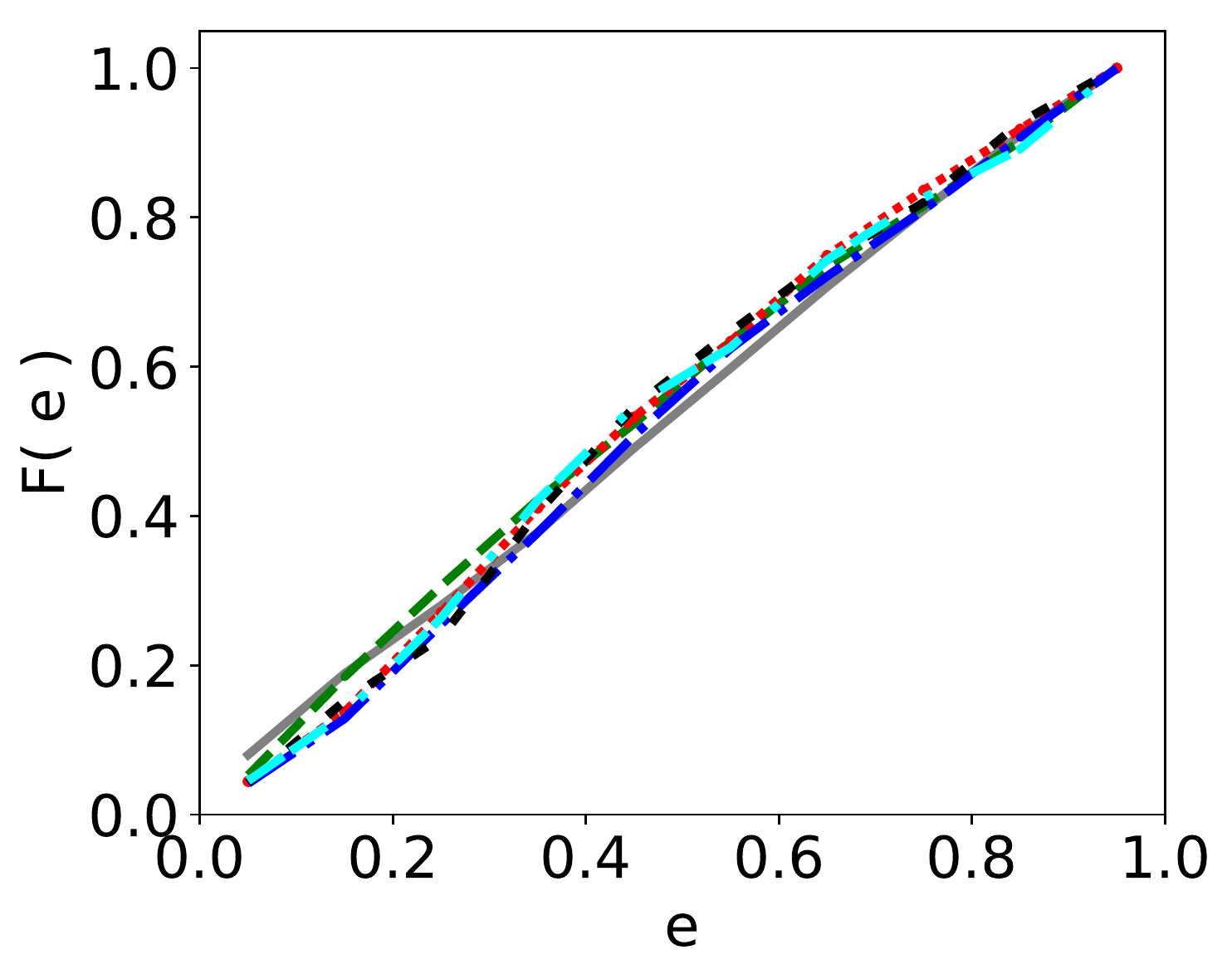}}
\end{minipage}
\caption{Cumulative distribution function of binary S-stars semi-major axis (left) and eccentricity (right) as function of $\ain$ for Model 1. More than $\approx 90$\% of binary S-stars have $a_b\lesssim 1$ AU to avoid tidal disruption by the MBH. The cumulative distribution of eccentricities is nearly independent on the initial $\ain$ and is $\propto e$, i.e. the eccentricity distribution is constant.}
\label{fig:binbh}
\end{figure*} 
As discussed previously, the energy change $\delta E$ imparted to the inner binary is several times larger than $E_b$ and, as a consequence, the binary semi-major axis can change significantly even if a small fraction of $\delta E$ is converted into binary internal energy. Figure \ref{fig:binbh} shows the cumulative distribution function of binary S-stars semi-major axis and eccentricity as a function of $\ain$ for Model 1.  There is a clear correlation between the final and initial binary semi-major axis. Also, more than $\approx 90\%$ of binary S-stars have $a_b\lesssim 0.5$ AU to avoid tidal disruption by the MBH. The cumulative distribution of eccentricities is nearly independent of the initial $\ain$ and is $\propto e$, i.e. the eccentricity distribution is constant.

\section{Discussions and Conclusions}

The recent discovery of a candidate HVB $\approx 5.7$ kpc away from the Galactic Centre and  travelling at $\sim 570\kms$ \citep{nem16} has 
brought new attention to the production of such fast-moving binaries. Moreover, the birthplace and ejection mechanism of the $\sim 9\msun$ main-sequence HVS HE0437-543 remain uncertain. WIth a main-sequence lifetime shorter than the flight time from the GC, an ejection from the GC due to a Hills type tidal disruption is ruled out. Possible scenarios include an origin in the Large Magellanic Cloud and ejection due to an encounter with a massive black hole \citep{GPZ2007}  and a blue straggler formation following a triple disruption in the GC and ejection of an HVB \citep{per09}. A similar scenario was proposed by \citet{fck17}, who showed that $\sim 7\%$ of the HVBs ejected by a compact young star cluster merge originating blue-straggler HVSs. However, currently available \textit{Hubble Space Telescope} proper motions for the star fail to constrain its birth location, with both a Milky Way and a Large Magellanic Cloud origin consistent with the measurements. As discussed in \citet{ede05}, the proper motion of HE0437-543 should be $\approx 0.01$ mas yr$^{-1}$ if it was produced in the GC, while $\approx 2$ mas yr$^{-1}$ if it originated in the LMC \citep{bou2017}. Proper motions from the \textit{Gaia} mission should constrain the star's origin in the near future.

\citet{per09} suggest that HVBs may be ejected following triple disruptions by the MBH in the GC. Some of these binaries may evolve into blue straggler stars as a result of their internal evolution after ejection, thereby resulting in a rejuvenation and a shorter apparent main-sequence lifetime.  In this work, we test the triple tidal disruption scenario by means of high-accuracy scattering experiments involving a triple star and the Milky Way's central MBH. We vary the inner and outer binary initial separation as well as the masses of the stars to include the most promising values suggested by \citet{per09}, and perform simulations with both point masses and finite stellar radii. We find that only a very small fraction ($\lesssim 1\%$) of encounters result in tidal disruption of the outer binary and ejection of the inner binary, for all sets of parameters. We explain this result with a simple theoretical argument based on the energy change produced in the encounter. Only three of the possible outcomes have significant probabilities, namely the ejection of a single HVS and the production of two single bound stars (1SH-2SS), the production of three single S-stars (3SS) and the production of one single and one binary S-star bound to the MBH (1SS-1BS).

To convert the $1\%$ probability of HVB production into a detection rate, we note that the 
typical plunge rate of tidal disruptions is of the order of the dynamical period at the influence radius. If we assume that nearly $13\%$ of stars are found in triples \citep{tok14b}, the typical rate of HVB production can be roughly estimated as $\approx 1\gyr^{-1}$. This translates into about 10 HVBs to be found in the Galaxy from the triple disruption mechanism. We note that this is probably an upper limit since the initial choice of circular inner and outer orbits in the triple probably limits the number of collisions. Moreover, also the choice of the triple stars fraction in the Galactic Center and the initial inner and outer semi-major axis is quite optimistic. For the former, we assumed the same fraction of the solar neighborhood, but it could be probably smaller in analogy with binary stars \citep{hop09}, while for the latter we assumed tight inner and outer orbits, which should be the more likely progenitors of the HVBs because of a larger energy reservoir. We conclude that triple disruptions in the GC followed by blue straggler formation are an unlikely source of HVBs in the Galaxy and alternative mechanisms need to be invoked to explain the origin of HVBs and HVSs with short lifetimes. These include encounters of stellar binaries with a massive black hole binary \citep{sesa2006,baum2006,lu2007}.

\section{Acknowledgements}
GF acknowledges hospitality from Mark Gieles and the University of Surrey, where the early plan for this work was conceived. GF thanks Seppo Mikkola for helpful discussions on the use of the code \textsc{ARCHAIN}. Simulations were run on the \textit{Astric} cluster at the Hebrew University of Jerusalem.

\bibliographystyle{mn2e}
\bibliography{biblio}
\end{document}